\documentclass [aps, pra, amssymb, twocolumn,superscriptaddress]{revtex4-1}
\usepackage{graphicx,epstopdf, amsfonts, amssymb, amsmath}
\usepackage{bm}
\usepackage{times}
\usepackage[colorlinks,citecolor=blue,linkcolor=red]{hyperref}
\usepackage{color}

\begin{document}

\title{Simulating Quantum Spin Hall Effect in Topological Lieb Lattice of Linear Circuit Network}

\author{Weiwei Zhu}
\thanks{These authors have contributed equally to this work.}
\affiliation{School of Physics Science and Engineering, Tongji University, 200092 Shanghai, P. R. China}
\author{Shanshan Hou}
\thanks{These authors have contributed equally to this work.}
\affiliation{School of Physics Science and Engineering, Tongji University, 200092 Shanghai, P. R. China}
\author{Yang Long}
\affiliation{School of Physics Science and Engineering, Tongji University, 200092 Shanghai, P. R. China}
\affiliation{%
Center for Phononics and Thermal Energy Science, China-EU Joint Center for Nanophononics, Shanghai Key Laboratory of Special Artificial Microstructure Materials and Technology,
School of Physics Sciences and Engineering, Tongji University, 200092 Shanghai, China
}%
\author{Hong Chen}
\affiliation{School of Physics Science and Engineering, Tongji University, 200092
Shanghai, P. R. China}
\affiliation{%
Center for Phononics and Thermal Energy Science, China-EU Joint Center for Nanophononics, Shanghai Key Laboratory of Special Artificial Microstructure Materials and Technology,
School of Physics Sciences and Engineering, Tongji University, 200092 Shanghai, China
}%
\author {Jie Ren}
\email{Xonics@tongji.edu.cn}
\affiliation{School of Physics Science and Engineering, Tongji University, 200092
Shanghai, P. R. China}
\affiliation{%
Center for Phononics and Thermal Energy Science, China-EU Joint Center for Nanophononics, Shanghai Key Laboratory of Special Artificial Microstructure Materials and Technology,
School of Physics Sciences and Engineering, Tongji University, 200092 Shanghai, China
}%


\begin{abstract}
Inspired by the topological insulator circuit experimentally proposed by Jia \textit{et al}. [Phys. Rev. X \textbf{5}, 021031 (2015)], we theoretically realize the topological Lieb lattice, a line centered square lattice with rich topological properties, in a radio-frequency circuit. We design a specific capacitor-inductor connection to resemble the intrinsic spin-orbit coupling, and construct the analog spin by mixing degrees of freedom of voltages. As such, we are able to simulate the quantum spin Hall effect in the topological Lieb lattice of linear circuits.
We then investigate the spin-resolved topological edge mode and the topological phase transition of band structure varied with capacitances.
Finally, we discuss the extension of $\pi/2$ phase change of hopping between sites to arbitrary phase values. Our results may find implications in engineering microwave topological metamaterials for signal transmission and energy harvesting.
\end{abstract}







\maketitle

\section{Introduction}

Topological insulator~\cite{TI1,TI2,TI3} is a new electronic quantum spin Hall state (QSHE) in condensed matter physics, which is insulating in the bulk but   conducting on the surface, with the conducting edge mode robustly protected by the band topology.
Later on, similar topological properties have been extended to
other kinds of artificial structure systems such as phononic~\cite{phonon1,phonon2,phonon3,phonon4} and photonic metamaterials and crystals~\cite{photon1,photon2,photon3,photon4,photon5,photon6,photon7,photon8}. The topological
phase of two-dimensional (2D) lattices can be distinguished by the topological invariants,
e.g., integer Chern number, which is the integration of the Berry curvature over 2D first Brillouin zone (BZ). As such, the lattice geometry plays the key role for the band topology.

\begin{figure}
\hspace{-2mm}
\scalebox{0.33}[0.33]{\includegraphics{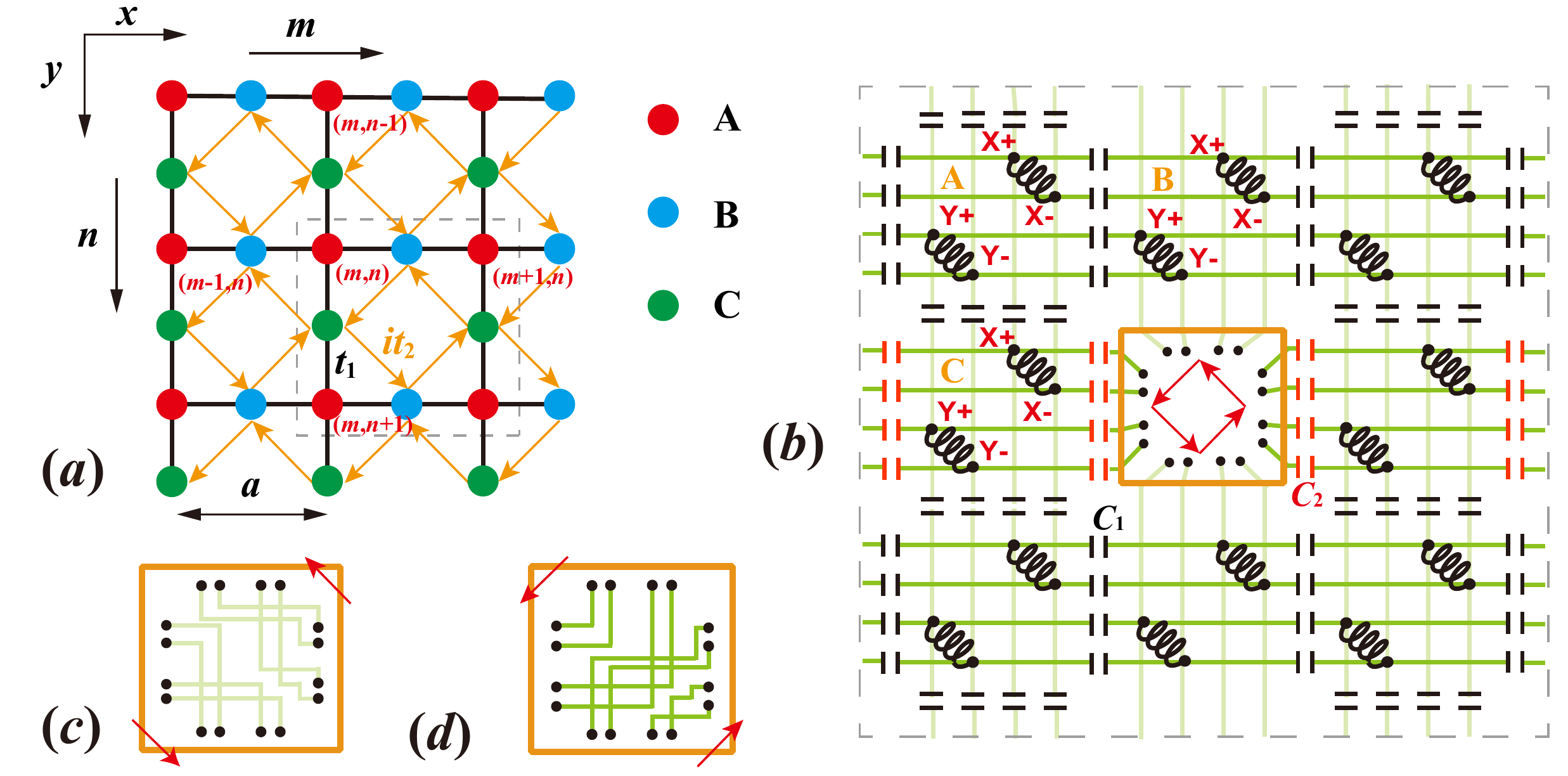}}
\vspace{-2mm}
\caption{(a) The schematic diagram of the topological Lieb lattice (TLL) with $9$ 
unit cells. Each unit cell is composed of three sites A, B, and C with lattice constant $a$ and is numbered by $(m,n)$. Nearest-neighbor (NN) interaction with strength $t_1$ is indicated by black line, and the intrinsic spin-orbit couplings (SOCs) with strength $t_2$ are indicted by orange arrows. (b) The designed linear circuit (LC) to mimic the dashed boxed part of TLL in (a). Each site is formed by two inductors, $X$ with two nodes $X_+$,$X_-$ and $Y$ with two nodes $Y_+$, $Y_-$. The NN interaction is achieved by the black coupling capacitors with capacitance $C_1$ and the SOCs are achieved by the red coupling capacitors with capacitance $C_2$, whose interconnects are shown in (c) and (d). And the interconnects shown in (c) and (d) have to be simultaneously implemented in the circuit to introduce intrinsic SOCs.}
\label{fig1}
\end{figure}

Among tons of lattice structures, the Lieb lattice exhibits intriguing properties that are of general interest in both the fundamental physics and practical applications.
The topological Lieb lattice, shown in Fig.~\ref{fig1}($a$), is the line centered square lattice with three sites A, B and C per unit cell. It is essentially a 2D counterpart of the perovskite structure~\cite{2,3,4}, which is ubiquitous in nature. Due to the peculiar lattice structure with spin-orbit couplings(SOCs) [see arrows in Fig.~\ref{fig1}($a$)]  for electronic tight-binding model, this lattice is characterized by three energy bands~\cite{5} and displays unusual topological properties, for instance, the protected flat band~\cite{16,8,9} and novel topological phase transition~\cite{6,17,He}.

For the infinite Lieb lattice without SOCs, the three bands touch each other at the middle of the spectrum, known as the zero energy, and exhibit a Dirac cone at the point $(\pi,\pi)$ in the first BZ with a flat band crossing that Dirac point, a so-called spin-1 Dirac cone.
People can  open the three-fold degeneracy into a multi-gapped structure by either adding a dimerization term that staggers the hopping along the x- and y-directions or by including a Rashba SOCs. But, the resulting gaps are topologically trivial, as was discussed by Weeks and Franz~\cite{2}. Nevertheless, the authors showed that an intrinsic SOCs, acting as a pure imaginary next-nearest-neighbor hooping, is able to separate the three bands by inducing topologically non-trivial band gaps. In this work, we will design and implement the intrinsic SOC in Lieb lattice [see arrows in Fig.~\ref{fig1}($a$)] using capacitor-inductor network of linear circuit (LC)~\cite{photon7,photon8,26}. We refer to this Lieb lattice with SOC as {\it topological Lieb lattice} (TLL) in this work.

So far, the Lieb lattice has been realized and researched in different physical platforms, e.g., the optical waveguides~\cite{7,8,9} by the direct laser writing technique~\cite{13} and cold atoms in optical lattices~\cite{10,11,12}. Later, by introducing the Floquet mechanism, the photonic TLL is designed with the synthetic SOCs formed by helical waveguides \cite{YL}. Although cold atoms and optical waveguides offer a high flexible control over lattice geometries and allow for the addition of synthetic and tunable interactions, they have strict requirements for experimental equipments and are applied specifically to waves of high frequency and short wavelength.

In this paper, we further develop the approach presented in Ref.~\cite{photon7} for Hofstadter model~\cite{14,15} to spin Lieb lattice in a simple and systematic way. We design a specific capacitor-inductor network to resemble SOCs, and construct the analog spin by mixing the degrees of freedom(DOFs) of voltages.
Meanwhile, the hopping phase between sites($\phi=\pi/2$) is also extended to arbitrary values. Therefore,  in the TLL of LCs we are able to simulate the QSHE characterized by nonzero spin Chern number, and to investigate the spin-resolved topological edge mode, as well as the topological phase transition by varying the capacitances. 

\section{Model and Results}
\subsection{Fabrication of Topological Lieb Lattice circuit}

We start from the electronic Hamiltonian of the tight-binding model for a 2D TLL with SOCs,
\begin{eqnarray}
  \hat{H} &=& t\sum_{i;s}\hat{c}^{\dagger}_{i,s}\hat{c}_{i,s} +t_1\sum_{\langle{i,j}\rangle;s}\hat{c}^{\dagger}_{i,s}\hat{c}_{j,s} \nonumber \\
  &+&i s t_2 \sum_{\ll{i,j}\gg;s}(\mathbf{e}_{i,j}\cdot \mathbf{e}_{z})\hat{c}^{\dagger}_{i,s}\hat{c}_{j,s}.
\label{eq1}
\end{eqnarray}
Here, $\hat{c}_{i,s}$($\hat{c}^{\dagger}_{i,s}$) is the annihilation(creation) operator for particle on site $i$ with spin $s$. The first term describes onsite potential and the second is the spin-independent nearest-neighbor (NN) hopping process of a particle with spin projection $s$ from site $i$ to site $j$. The third term describes the SOC term with coupling constant $t_2$ [see Fig.~\ref{fig1}($a$)]. $\mathbf{e}_{z}$ denotes the unit vector in the z-direction. Moreover, we define for the NN SOC a unit vector $\mathbf{e}_{i,j}=\mathbf{e}_{i,k}\times \mathbf{e}_{k,j}/\mid \mathbf{e}_{i,k}\times \mathbf{e}_{k,j}\mid$ in terms of the bond vectors $\mathbf{e}_{i,k}$ and $\mathbf{e}_{k,j}$ that connects the sites $i$ and $j$ via the unique intermediate site $k$. By applying a Fourier transformation to the real space Hamiltonian in Eq.(\ref{eq1}), we obtain the Hamiltonian in momentum space,

\begin{equation}\label{eq2}
  \hat{H}=\sum_{k\in BZ}\hat{\Psi}^{\dagger}_{k}H_{k}\hat{\Psi}_{k},  H_{k}=H_{k}^{0}\otimes I_{2\times2}+H_{k}^{\mathrm{SOC}}\otimes\sigma_{z},
\end{equation}
where $\hat{\Psi}_{k}\equiv(\hat{\Psi}_{k,\uparrow},\hat{\Psi}_{k,\downarrow})$ with $\hat{\Psi}_{k,s}\equiv(\hat{c}_{k,s}^{A},\hat{c}_{k,s}^{B},\hat{c}_{k,s}^{C})$ where $s=\uparrow,\downarrow$, and the $3\times3$ matrix $H_{k}^{0}$ and $H_{k}^{\mathrm{SOC}}$ are given by
\begin{equation}\label{eq3}
  H_{k}^{0}=\left(\begin{array}{ccc}
                    t & 2t_{1}\cos(\frac{k_{x}a}{2}) & 2t_{1}\cos(\frac{k_{y}a}{2}) \\
                    2t_{1}\cos(\frac{k_{x}a}{2}) & t & 0 \\
                     2t_{1}\cos(\frac{k_{y}a}{2}) & 0 & t
                  \end{array}
  \right),
\end{equation}
and
\begin{equation}\label{eq4}
\begin{split}
  &H_{k}^{\mathrm{SOC}}= \\
  &4it_{2}\left(\begin{array}{ccc}
                    0 & 0 &0 \\
                   0 & 0 & \sin(\frac{k_{x}a}{2})\sin(\frac{k_{y}a}{2}) \\
                    0 & -\sin(\frac{k_{x}a}{2})\sin(\frac{k_{y}a}{2}) & 0
                  \end{array}
  \right).
  \end{split}
\end{equation}
The Hamiltonian matrix $H_{k}$ consists of two decoupled blocks corresponding to the spin up and spin down projections, related by fermi time-reversal symmetry $H_{k,\uparrow}=H_{-k,\downarrow}^{*}$.

In order to simulate the TLL with SOC described by Eq.~(\ref{eq1}), as well as the QSHE, let us now describe the procedure of designing the corresponding LC analogy made of capacitor-inductor networks and constructing the pseudo-spin DOF. Indeed, there are a variety of ways~\cite{photon7,photon8,19,20,21,22,23} to engineer topologically non-trivial band structures in lattice models. However, there is no doubt that the LC made by lumped elements is the most convenient and simple method among them, which is around the radio-frequency (RF) regime and can be easily applied to the microwave engineering. Therefore, in the present work we realize the QSHE by designing TLL LC network with SOC and investigate the spin-resolved topological edge transport and topological phase transition. We inherit and extend the approach that two arrays of inductors provide the spin up and spin down RF photons presented in Ref.~\cite{photon7}. The specific connections of our LC model is shown in Figs.~\ref{fig1}($b$), ($c$) and ($d$). And the interconnects shown in both Fig.~\ref{fig1}($c$) and ($d$) have to be simultaneously implemented in the circuit to introduce the intrinsic SOCs.

Let us describe these connections (analogy) in detail. First, we define that plaquette A, B and C of Fig.~\ref{fig1}($b$) forms a unit and is mapped to the atomic site A, B and C of a unit in Fig.~\ref{fig1}($a$), respectively. Each site contains two inductors X and Y, which behave as two sub-orbitals. We fix respectively both ends of the inductors as $X_+$ and $X_-$ or $Y_+$ and $Y_-$. And we define the voltages across inductors X, Y : $U_X=V_{X_+}-V_{X_-}$ and $U_Y=V_{Y_+}-V_{Y_-}$ . As such, the positive or negative coupling is controlled by which ends of the on-site inductors are capacitively connected to one another. Therefore, the connections of X-to-X or Y-to-Y simulates the NN couplings, and the SOC can be generated through the connection X-to-Y in the central plaquette, if we define $U_{\uparrow}=U_X+i U_Y$ and $U_{\downarrow}=U_X-i U_Y$, respectively.
For instance, through the couplings indicated by solid arrows between plaquette B and plaquette C that are used to simulate the SOC, we can realize $U_X \to -U_Y$ and $U_Y \to U_X$, so that $U_{\uparrow} \to i U_{\uparrow}$ and $U_{\downarrow} \to -i U_{\downarrow}$.

To see more details, the dynamical properties of the system in one unit cell are described according to the Kirchhoff's law. The detailed derivation process is shown in Appendix A. Here we show the six equations of motion,
\begin{widetext}
\begin{equation}\label{eq5}
\left\{ \begin{array}{cc}
U_{m,n,X}^{A}=\frac{LC}{2}(-4t_1\ddot{U}_{m,n,X}^{A}+t_1(\ddot{U}_{m-1,n,X}^{B}+\ddot{U}_{m,n,X}^{B}+\ddot{U}_{m,n-1,X}^{C}+\ddot{U}_{m,n,X}^{C})) \\
U_{m,n,Y}^{A}=\frac{LC}{2}(-4t_1\ddot{U}_{m,n,Y}^{A}+t_1(\ddot{U}_{m-1,n,Y}^{B}+\ddot{U}_{m,n,Y}^{B}+\ddot{U}_{m,n-1,Y}^{C}+\ddot{U}_{m,n,Y}^{C})) \\
U_{m,n,X}^{B}=\frac{LC}{2}(-(2t_1+4t_2)\ddot{U}_{m,n,X}^{B}+t_2(\ddot{U}_{m+1,n,Y}^{C}-\ddot{U}_{m+1,n-1,Y}^{C}-\ddot{U}_{m,n,Y}^{C}+\ddot{U}_{m,n-1,Y}^{C})+t_1(\ddot{U}_{m,n,X}^{A}+\ddot{U}_{m+1,n,X}^{A})) \\
U_{m,n,Y}^{B}=\frac{LC}{2}(-(2t_1+4t_2)\ddot{U}_{m,n,Y}^{B}-t_2(\ddot{U}_{m+1,n,X}^{C}-\ddot{U}_{m+1,n-1,X}^{C}-\ddot{U}_{m,n,X}^{C}+\ddot{U}_{m,n-1,X}^{C})+t_1(\ddot{U}_{m,n,Y}^{A}+\ddot{U}_{m+1,n,Y}^{A})) \\
U_{m,n,X}^{C}=\frac{LC}{2}(-(2t_1+4t_2)\ddot{U}_{m,n,X}^{C}-t_2(\ddot{U}_{m,n+1,Y}^{B}-\ddot{U}_{m,n,Y}^{B}-\ddot{U}_{m-1,n+1,Y}^{B}+\ddot{U}_{m-1,n,Y}^{B})+t_1(\ddot{U}_{m,n,X}^{A}+\ddot{U}_{m,n+1,X}^{A})) \\
U_{m,n,Y}^{C}=\frac{LC}{2}(-(2t_1+4t_2)\ddot{U}_{m,n,Y}^{C}+t_2(\ddot{U}_{m,n+1,X}^{B}-\ddot{U}_{m,n,X}^{B}-\ddot{U}_{m-1,n+1,X}^{B}+\ddot{U}_{m-1,n,X}^{B})+t_1(\ddot{U}_{m,n,Y}^{A}+\ddot{U}_{m,n+1,Y}^{A}))
\end{array}\right.
\end{equation}
\end{widetext}
Here, $U_{m,n,i}^{j}$ is the voltage difference of inductor $i$ in plaquette $j$ of the $(m,n)^{th}$ unit cell with $i=X,Y$ and $j=A,B,C$. $L$ is the inductance. $C_1=t_1C$ and $C_2=t_2C$ denote capacitances connecting the sites A-to-B or A-to-C and B-to-C, respectively.
Since the local circuit dynamics for X and
Y are coupled with each other, in order to obtain the QSHE, we need to decouple the entangled Lieb LC into two copies of spinful Lieb LCs. As such, we can certainly describe the structure in the ``spin'' representation by denoting $U_{\uparrow,\downarrow}=U_X\pm iU_Y$, and then finally get two sets of decoupled equations for $U_{\uparrow}$ and $U_{\downarrow}$ respectively, exhibiting opposite effective gauge fields:
\begin{widetext}
\begin{equation}\label{eq6}
\left\{ \begin{array}{cc}
U_{m,n,\uparrow}^{A}=\frac{LC}{2}\left(-4t_1\ddot{U}_{m,n,\uparrow}^{A}+t_1(\ddot{U}_{m-1,n,\uparrow}^{B}+\ddot{U}_{m,n,\uparrow}^{B}+\ddot{U}_{m,n-1,\uparrow}^{C}+\ddot{U}_{m,n,\uparrow}^{C})\right) \\
U_{m,n,\uparrow}^{B}=\frac{LC}{2}\left(-(2t_1+4t_2)\ddot{U}_{m,n,\uparrow}^{B}-it_2(\ddot{U}_{m+1,n,\uparrow}^{C}-\ddot{U}_{m+1,n-1,\uparrow}^{C}-\ddot{U}_{m,n,\uparrow}^{C}+\ddot{U}_{m,n-1,\uparrow}^{C})+t_1(\ddot{U}_{m,n,\uparrow}^{A}+\ddot{U}_{m+1,n,\uparrow}^{A})\right) \\
U_{m,n,\uparrow}^{C}=\frac{LC}{2}\left(-(2t_1+4t_2)\ddot{U}_{m,n,\uparrow}^{C}+it_2(\ddot{U}_{m,n+1,\uparrow}^{B}-\ddot{U}_{m,n,\uparrow}^{B}-\ddot{U}_{m-1,n+1,\uparrow}^{B}+\ddot{U}_{m-1,n,\uparrow}^{B})+t_1(\ddot{U}_{m,n,\uparrow}^{A}+\ddot{U}_{m,n+1,\uparrow}^{A})\right) \\
U_{m,n,\downarrow}^{A}=\frac{LC}{2}\left(-4t_1\ddot{U}_{m,n,\downarrow}^{A}+t_1(\ddot{U}_{m-1,n,\downarrow}^{B}+\ddot{U}_{m,n,\downarrow}^{B}+\ddot{U}_{m,n-1,\downarrow}^{C}+\ddot{U}_{m,n,\downarrow}^{C})\right) \\
U_{m,n,\downarrow}^{B}=\frac{LC}{2}\left(-(2t_1+4t_2)\ddot{U}_{m,n,\downarrow}^{B}+it_2(\ddot{U}_{m+1,n,\downarrow}^{C}-\ddot{U}_{m+1,n-1,\downarrow}^{C}-\ddot{U}_{m,n,\downarrow}^{C}+\ddot{U}_{m,n-1,\downarrow}^{C})+t_1(\ddot{U}_{m,n,\downarrow}^{A}+\ddot{U}_{m+1,n,\downarrow}^{A})\right) \\
U_{m,n,\downarrow}^{C}=\frac{LC}{2}\left(-(2t_1+4t_2)\ddot{U}_{m,n,\downarrow}^{C}-it_2(\ddot{U}_{m,n+1,\downarrow}^{B}-\ddot{U}_{m,n,\downarrow}^{B}-\ddot{U}_{m-1,n+1,\downarrow}^{B}+\ddot{U}_{m-1,n,\downarrow}^{B})+t_1(\ddot{U}_{m,n,\downarrow}^{A}+\ddot{U}_{m,n+1,\downarrow}^{A})\right)
\end{array}\right.
\end{equation}
\end{widetext}
We can see clearly that there are two spin copies of the Lieb lattice model for $U_{\uparrow}$ and $U_{\downarrow}$ respectively. And the coupling coefficients between site B and site C is a pure imaginary number, which means an effective gauge field with $\pi/2$ coupling phase and simulates the SOCs. We can get the equation of motion for $U_{\downarrow}$ from the conjugate of the equation of motion for $U_{\uparrow}$. Thus the circuit model has opposite effective (magnetic) gauge field for $U_{\uparrow}$ and $U_{\downarrow}$. Choosing the dynamical factor $e^{i\omega t}$ and applying a Fourier transformation to the equations of motion for spin up and spin down in Eqs.(\ref{eq6}), we get the equations of motion in momentum space,

\begin{equation}\label{eq7}
  \frac{1}{\omega^{2}}U_{k}=M_{k}U_{k},M_{k}=M_{k}^{0}\otimes I_{2\times2}+M_{k}^{\mathrm{SOC}}\otimes\sigma_{z},
\end{equation}
where $U_{k}\equiv(U_{k,\uparrow},U_{k,\downarrow})^{T}$ with $U_{k,s}\equiv(U_{k,s}^{A},U_{k,s}^{B},U_{k,s}^{C})^{T}$, $T$ means the transpose of a matrix. The $3\times3$ matrix $M_{k}^{0}$ and $M_{k}^{\mathrm{SOC}}$ are given by
\begin{equation}\label{eq8}
\begin{split}
  &M_{k}^{0}=\frac{LC}{2}\times \\
  &\left(\begin{array}{ccc}
                    4t_{1} & -2t_{1}\cos(\frac{k_{x}a}{2}) & -2t_{1}\cos(\frac{k_{y}a}{2}) \\
                    -2t_{1}\cos(\frac{k_{x}a}{2}) & 2t_{1}+4t_{2} & 0 \\
                   -2t_{1}\cos(\frac{k_{y}a}{2}) & 0 & 2t_{1}+4t_{2}
                  \end{array}
  \right),
\end{split}
\end{equation}

and
\begin{equation}\label{eq9}
\begin{split}
  &M_{k}^{\mathrm{SOC}}=2it_{2}LC\times \\
  &\left(\begin{array}{ccc}
           0 & 0 & 0 \\
           0 & 0 & -\sin(\frac{k_{x}a}{2})\sin(\frac{k_{y}a}{2}) \\
           0 & \sin(\frac{k_{x}a}{2})\sin(\frac{k_{y}a}{2}) & 0
         \end{array}
  \right).
\end{split}
\end{equation}
The matrix $M_{k}^{0}$ and $M_{k}^{\mathrm{SOC}}$ have the same mathematical forms with $H_{k}^{0}$ and $H_{k}^{\mathrm{SOC}}$ in Eq.(\ref{eq3}) and Eq.(\ref{eq4}), which confirms that we can use the TLL LC network to investigate the spin-resolved topological edge transport and topological phase transition. Here we note that although $M_{k}$ has the eigenvalue $\frac{1}{\omega^{2}}$ instead of conventional $\omega^{2}$ ($\omega$ for electron systems) we still can get a topological nontrivial phase which is mainly determined by the eigenvectors.
\begin{figure*}
\hspace{-2mm}
\scalebox{0.3}[0.3]{\includegraphics{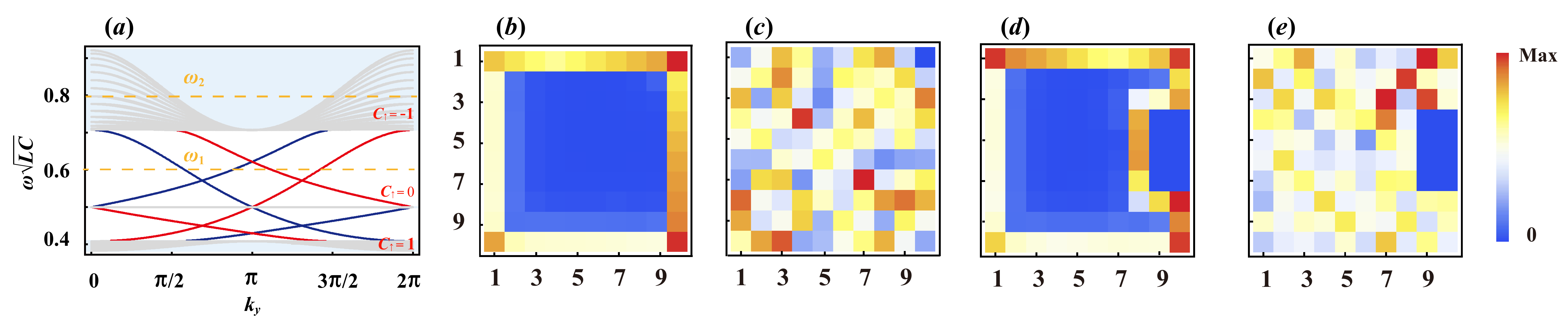}}
\vspace{-2mm}
\caption{($a$) Band structure of the semi-infinite TLL LC with capacitors $C_1/C_2=2$, i.e., a wide strip of circuit with fixed boundary in the transverse $x$-direction and periodic boundary conditions in the longitudinal $y$-direction, including massive bulk bands (gray) and spin-orbit-locked edge states ($U_{\uparrow}$ in red, $U_{\downarrow}$ in blue) that reside in the bulk gap. The Chern numbers of the spin-up bands are indicated next to each band. ($b$) Numerical calculation results of field strength with $10\times10$ units. The field strength for each unit cell is defined as $U_{m,n}=\sum_{i,j}|U_{m,n,i}^{j}|$ with $i=X,Y$, $j=A,B,C$. The excitation frequency $\omega_{1}$ is marked as orange line between upper and middle bands in (a). {\color{blue} Note the right and bottom sides of the finite system are terminated with B sites and C sites, respectively, as indicated in Fig.~\ref{fig1}(a). Thus the SOC interconnection parts are halved at those sides and the finite system lacks the inversion symmetry, so that the edge state does not have a symmetric field distribution.} ($d$) shows the field strength distribution with a defective system and the excitation frequency is the same as ($b$). ($c$) and ($e$) are the corresponding field strength distribution driven by the microwave of frequency $\omega_{2}$ marked in the upper band in (a).}
\label{fig2}
\end{figure*}

For electron systems, the QSHE is protected by the fermionic time-reversal symmetry $T_{f}=i\sigma_{y}K$ ($T^2_{f}=-1$), resulting in the Kramers doublet of spin-up and spin-down states. Unlike electrons, the second quantization of the voltage vibrations in RF microwave systems leads to photons that are massless bosons with spin-1. As such, conventional photonic system does not hold the Kramers degeneracy, but keeps the bosonic time-reversal symmetry $T_{b}=\sigma_xK$  ($T^2_{b}=1$). Therefore, we need introduce more DOFs to construct a pseudospin to satisfy Kramers degeneracy with fermionic-like pseudo time-reversal symmetry $T_{f}=i\sigma_{y}K$ ($T^2_{f}=-1$). In our system we double the DOFs by introducing two sub-orbitals
 (X and Y inductors) to each sites. By transforming into the pseudo spin basis, the constructed matrix $M_{k}$  does not only keep the bosonic time-reversal symmetry $T_{b}=I_{3\times3}\otimes\sigma_{x}K$ with $[T_{b},M_{k}]=0$, but also  further possesses the ferminoic-like  pseudo time-reversal symmetry $T_{f}=I_{3\times3}\otimes i\sigma_{y}K$ with $[T_{f},M_{k}]=0$. This fermionic pseudo time-reversal symmetry guarantees the Kramers degeneracy that leads to the topologically protected edge state transport. The topological photonic state with pseudo time-reversal symmetry but broken bosonic time-reversal symmetry has also been discussed~\cite{Lu}.

\subsection{Topological Band of the TLL LC  Model and QSHE}

Based on previous researches~\cite{2,6}, we understand that the effect of intrinsic SOCs is to open non-trivial gaps at the point $(\pi,\pi)$, i.e., when $t_2=0$ there is always a single spin-1 Dirac-cone gapless dispersion touching a flat band at the edge of BZ. Here we show that we can get the topologically non-trivial gaps, manifested as the topological edge state in a finite structure, in the TLL LC models. To demonstrate the topological edge state and QSHE, we get the band structure of an infinite TLL strip in y-direction with longitudinal quasi-momentum $k_y$, and finite in the transverse x-direction with 15 units. The band structure is shown in Fig.~\ref{fig2}($a$). Three bulk bands (gray curves) correspond to the bulk response of the system. Spin-helicity-coupled edge channels, as characteristics of a topological band structure, occupy the gaps between bulk bands. We can see that different spin excitations generate different propagation directions and can be ascertained from the slope of the energy-momentum dispersion. The edge states of opposite spins exhibit the symmetry of $(k,\uparrow) \leftrightarrow (-k,\downarrow)$, which is guaranteed by the pseudospin time-reversal symmetry. Additionally, the topological character of each spin-resolved band can be formally characterized in terms of a spin-Chern number $C_{\uparrow}$($C_{\downarrow}$), for $U_{\uparrow}(U_{\downarrow})$, which is $-1(+1)$ for the top band, $+1(-1)$for the bottom band and 0 for the middle flat band. This is consistent with the tight-binding model.

To probe this topological physics directly, numerical simulations are performed to demonstrate the topologically protected boundary modes on a finite TLL structure with $10 \times 10$ cell. We simulate the field strength distribution  of all sites at different frequencies marked by orange lines on Fig.~\ref{fig2}($a$). The results are shown in Fig.~\ref{fig2}($b$) and ($c$), clearly demonstrating the edge modes and bulk modes, respectively.  Afterward, for the sake of showing the stability of these edge modes, we artificially introduce defects at the right edge of the system, and the corresponding field strength distributions are shown in Fig.~\ref{fig2}($d$) and ($e$). One can clearly observe that the edge modes without back-scattering are not affected by the imperfections. Therefore, we can safely draw the conclusion that the edge modes are protected by nontrivial band topology. These phenomena confirm the existence of edge modes in QSHE and strongly suggest their topological nature.

\begin{figure*}
  \centering
  \includegraphics[width=5.2 in]{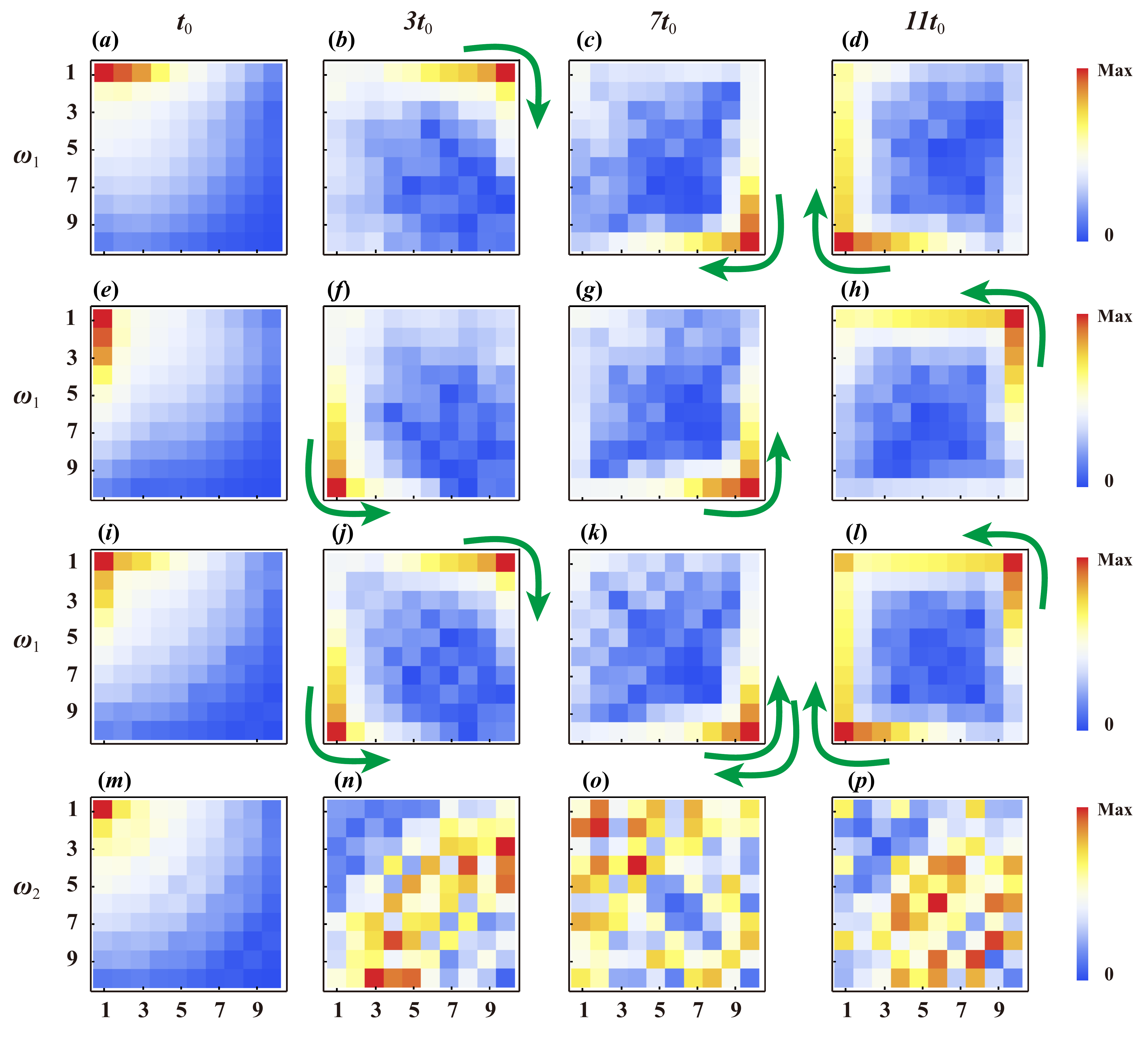}\\
  \caption{{The time evolution of Gaussian wave packet.} Four plots of each row indicate the trajectory of wave packet in turn and arrows indicate the direction of movement. The first three rows show the results excited by $\omega_1=0.6/\sqrt{LC}$ in the non-trivial gap, and the last one is $\omega_2=0.8/\sqrt{LC}$ inside the bulk band. The color bar denotes the strength of the field distribution. When calculating, we apply capacitors $C_3=t_3C$ to link A, B and C to the ground, which is used to compensate for different onsite potential due to some missing couplings at boundary.}
  \label{fig3}
\end{figure*}

In addition, we also implement a numerical simulation in time domain with a Gaussian wave packet source to see the time evolution of the wave packet. The simulation is carried with fourth order Runge-Kutta method. The simulation step length is less than $\frac{2\pi}{100\omega_{1}}$. The simulation results are shown in Fig.\ref{fig3}. When the corner site is driven by the sources that $S_X=e^{-\frac{(t-t_{0})^{2}}{\sigma^2}}\cos(\omega_1 t)$ and $S_Y=e^{-\frac{(t-t_{0})^{2}}{\sigma^2}}\sin(\omega_1 t)$ on the inductor X and Y of site A, respectively, we can see the wave packet spreads along the boundary clockwise as shown in Fig.\ref{fig3}(a)(b)(c)(d). At time $t_{0}$, the Gaussian wave peak enter the system.  Due to $S_{\uparrow}=S_X+iS_Y$, it is equivalent to excite the spin-up state $U_{\uparrow}=U_X+i U_Y$ alone with the spin-up source $S_{\uparrow}=e^{-\frac{(t-t_{0})^{2}}{\sigma^2}}e^{i\omega_{1} t}$. Similarly, we can solely drive the spin-down state $U_{\downarrow}=U_X-i U_Y$ by the spin-down source $S_{\downarrow}=S_X-iS_Y$ by setting $S_X=e^{-\frac{(t-t_{0})^{2}}{\sigma^2}}\cos(\omega_1 t)$ and $S_Y=-e^{-\frac{(t-t_{0})^{2}}{\sigma^2}}\sin(\omega_1 t)$, and we can observe the wave packet's counter-propagation as shown in Fig.\ref{fig3}(e)(f)(g)(h). If the driven voltage is $S_Y=0$ and $S_X=2e^{-\frac{(t-t_{0})^{2}}{\sigma^2}}\cos(\omega_1 t)=S_{\uparrow}+S_{\downarrow}$, the initial excitation contains both spin-up and spin-down source states, which will excite states to propagate in opposite directions along the boundary, behaving like the two opposite spin transports in QSHE as shown in Fig.\ref{fig3}(i)(j)(k)(l). When we change the frequency to the bulk band with frequency $\omega_{2}$ with source $S_X=2e^{-\frac{(t-t_{0})^{2}}{\sigma^2}}\cos(\omega_2 t)$ and $S_Y=0$, bulk mode excitations just trivially spread throughout the system as shown in Fig.\ref{fig3}(m)(n)(o)(p).

In Fig.\ref{fig3}, we see that the bulk states are also excited for frequency $\omega_{1}$. There are two reasons. Firstly, the sample is very small that is not exactly described by the band structure of infinite periodic systems, and the edge states can couple with each other through the wave function overlap of their decay wave into the bulk. Secondly, to see the wave dynamic evolution, we need the wave is localized enough so we choose the Gaussian wave packet with $S(t)=e^{-\frac{(t-t_{0})^{2}}{\sigma^2}}e^{i\omega_{1} t}$. Through a Fourier transformation to the frequency domain,
$S(t)=\frac{1}{\sqrt{2\pi}}\int_{-\infty}^{\infty}S(\omega)e^{i\omega t}d\omega,$
 we get,
\begin{equation}\label{eq11}
S(\omega)=\frac{\sigma}{\sqrt{2}}e^{-\frac{(\omega-\omega_{1})^{2}}{4/\sigma^{2}}}e^{-i(\omega-\omega_{1})t_{0}}.
\end{equation}
We see that the mode with frequency $\omega_{1}$ is the main component, but there are also other mixed modes, which may locate within the bulk bands. In our calculation we choose $\sigma=50$, $t_{0}=50\sqrt{LC}$.  Simulation details are shown in Appendix C.

\subsection{Topological phase transition of QSHE in the TLL}

Now, we discuss the effect of capacitances on the band structure, and study the topological phase transition driven by tuning the capacitance ratio $C_1/C_2$ in the presence of SOC. In other words, we are going to demonstrate topological phase transitions of QSHE  in our LC system by varying some capacitance.

\begin{figure}
\hspace{-2mm}
\scalebox{0.3}[0.3]{\includegraphics{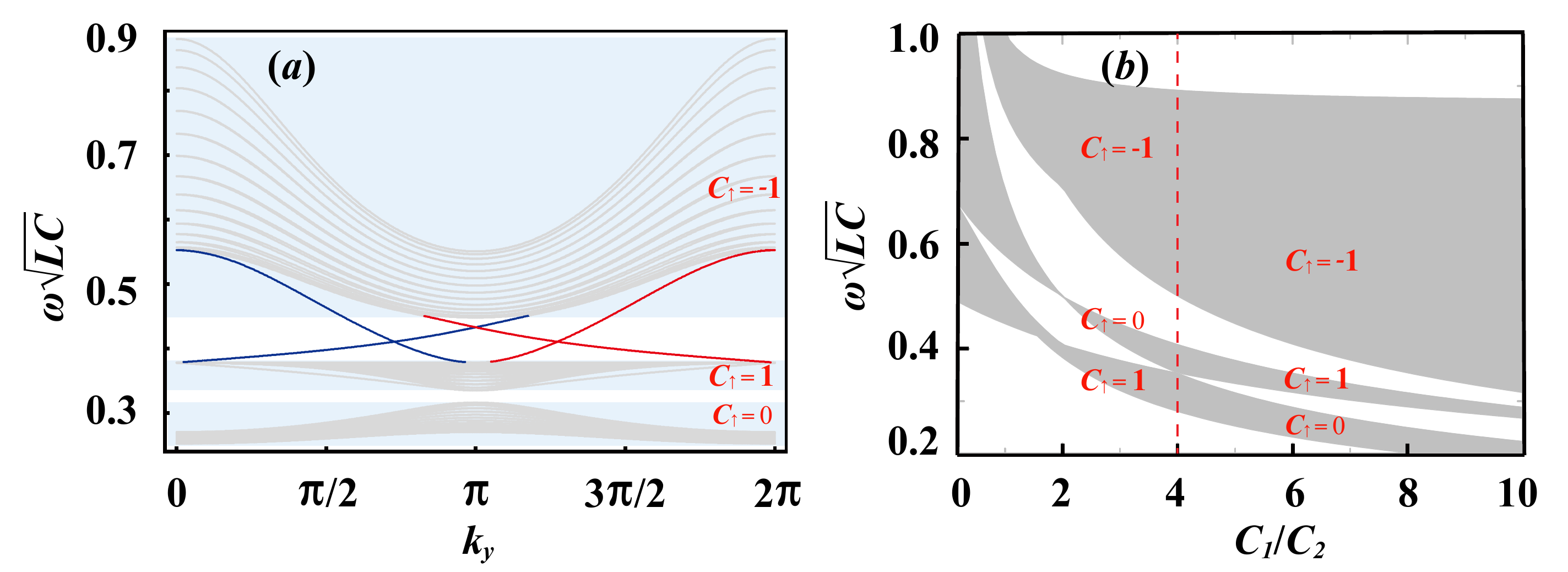}}
\vspace{-2mm}
\caption{($a$) Band structure of the semi-infinite TLL circuit with coupling capacitors $C_1/C_2=5$. Spin-orbit-locked edge states ($U_{\uparrow}$ in red, $U_{\downarrow}$ in blue) reside in the bulk gap. The Chern numbers of the spin-up bands are indicated next to each band. ($b$) Phase diagram as a function of $C_{1}/C_{2}=t_1/t_2$, and topological phase transition at a critical point $C_{1}/C_{2}=4$ where the second band gap closes. The spin up Chern numbers are marked by red letters.}
\label{fig4}
\end{figure}

We vary $C_1$ while keep $C_2$ fixed and observe the effect on the band structure by computing the spin Chern number. For the convenience of comparison, we focus on calculating Chern number for spin up state $C_{\uparrow}$. Fig.~\ref{fig4}($b$) illustrates the spin Chern number phase diagram of the system as a function of $C_1/C_2$, namely, the ratio of $t_1/t_2$. We can see that there are two regions in the phase diagram, with the critical point at $t_1/t_2=4$, where the spin-up Chern number change abruptly. Indeed, the change in spinful Chern number reflects the topology change of their respective spinful bands. In other words, middle and lower bands degenerate to close their gap when $t_1/t_2=4$, and then reopen as increasing the ratio.

Before the degeneracy, the energy spectrum diagram is just Fig.~\ref{fig2}($a$), with nontrivial topological spin edge mode in the lower gap; while after the gap reopening, the band structure is shown in Fig.~\ref{fig4}($a$), with the topological edge mode absent in the lower gap. Accordingly, the spin-up Chern numbers of three bands $(C_{up},C_{mid},C_{down})$ change from $(-1,0,1)$ to $(-1,1,0)$. In these calculations, the capacitor $C_2$ connecting sites B-to-C is fixed, while the capacitor $C_1$ is varied. These topological phase transitions of QSHE give a way to modify the transport behavior of the TLL LC at a particular frequency from unidirectional propagation along the edge to propagation-forbidden along the edges and to the interior.

\subsection{Generalization to arbitrary phase}

\begin{figure}
\hspace{-1mm}
\scalebox{0.4}[0.4]{\includegraphics{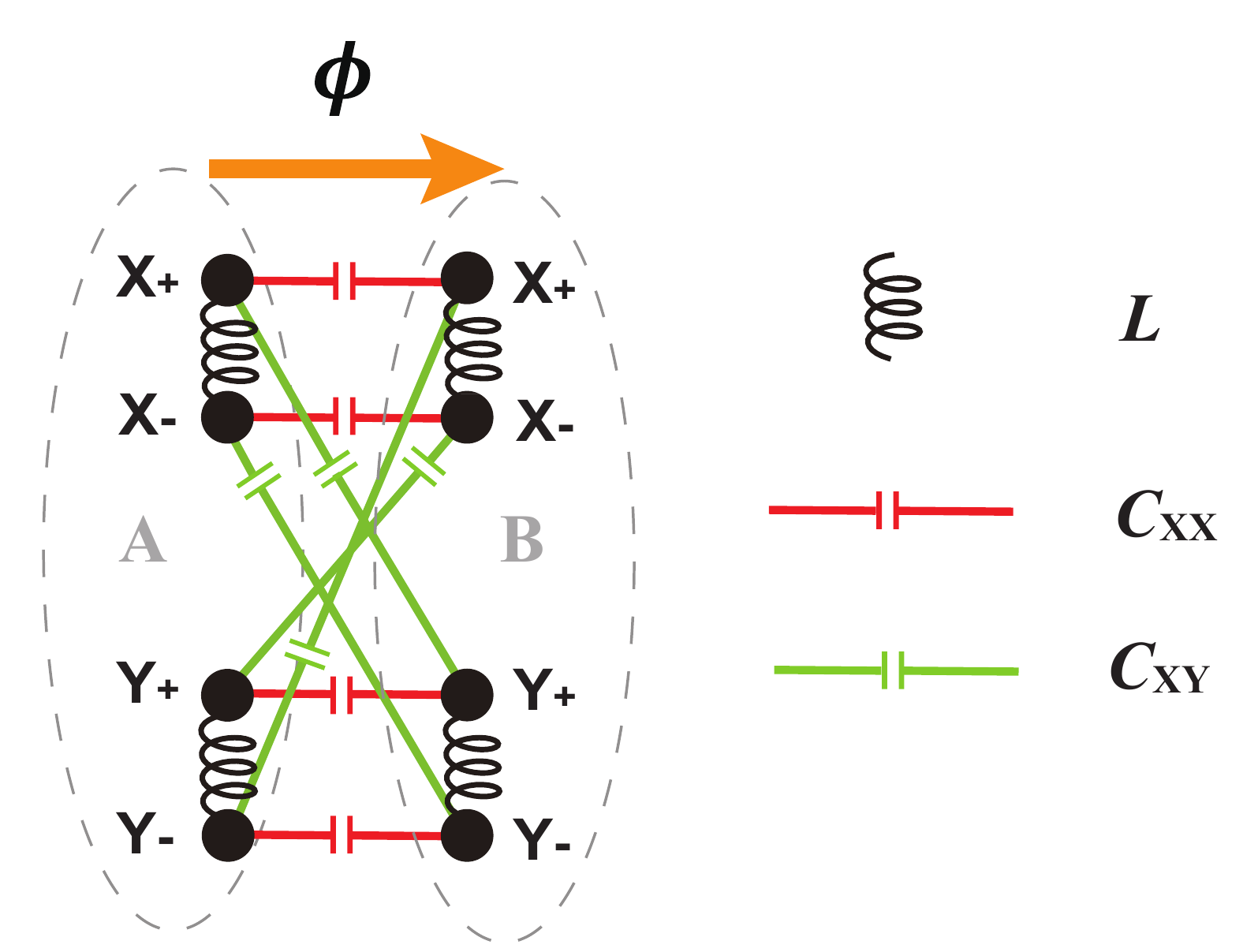}}
\vspace{-1mm}
\caption{The schematic diagram of realizing arbitrarily phase change $\phi$ between sites $A$ and $B$. The direction of the light brown arrow represents the orientation of phase change. The structure is formed by inductors (black spring) and coupling capacitors (two vertical short line) that are connected via wires (green and red lines). $X_+,X_-,Y_+$ and $Y_-$ indicate end of inductors. The value of the capacitance for capacitor and the inductance for inductor are marked on the right. The phase $\phi$ is related with the capacitances by $\phi=\tan^{-1}(\frac{C_{XY}}{C_{XX}})$.}
\label{fig5}
\end{figure}
So far, our above design merely has four connections between next-nearest-neighbor sites, which can  realize the only gauge field $e^{i\pi/2}$ for the SOC. One may consider increasing the internal number of connections and generalizing the gauge field to arbitrary phase. Actually, the arbitrary phase coupling has been proposed by V. Albert \textit{et al}~\cite{photon8}. They use the capacitors as sites, which are connected by inductors while in present work we do the duality case where the sites are composed of inductors connected by capacitors. Therefore, in V. Albert \textit{et al}'s work~\cite{photon8}, the phase is tuned by the ratio of inductances, while as will see in our case, the arbitrary phase is tuned by the ratio of capacitances.

As shown in Fig.~\ref{fig5}, the new designed structure can be formed by inductors (black spring) and coupling capacitors (two vertical short line). And different colors denote different capacitances. The definition of other parameters are same as the previous. This specialized connections can simulate hopping matrices with arbitrarily background magnetic flux. The corresponding rotation matrix from the left site voltage to the right site voltage reads $C\left(\begin{array}{ccc}
\cos\phi & \sin\phi \\
-\sin\phi & \cos\phi
\end{array} \right)$, where $C=\sqrt{C_{XX}^{2}+C_{XY}^{2}}$ and the phase
\begin{equation}
\phi=\tan^{-1}(\frac{C_{XY}}{C_{XX}}).
\end{equation}
That is, the horizontal capacitive connections of both X-to-X and Y-to-Y provide $C\cos\phi$ and the crossing capacitive connections of X-to-Y and Y-to-X provide $C\sin\phi$. When the end $\pm$ connects to the end $\pm$ the coupling is positive, while that the end $\pm$ connects to the end $\mp$ leads to the negative coupling since the voltage is reversed.
Through those capacitive connections, we can realize $U_X \to U_XC\cos\phi-U_YC\sin\phi$ and $U_Y \to U_Y C\cos\phi+ U_X C\sin\phi$, i.e. $U_{\uparrow} \to Ce^{i \phi}U_{\uparrow}$ and $U_{\downarrow} \to Ce^{-i \phi}U_{\downarrow}$. A detailed derivation process for the realization of arbitrary phase is shown in Appendix D.

We note that for the present configuration, the positive capacitance values restrict the value of $\phi$ in the range of [$0,{\pi}/{2}$], which also leads to [$\pi,{3\pi}/{2}$] when we exchange the sites $A$ and $B$. Based on the two configurations, if we simultaneously change the connections ($X_{\pm}^{A}$ to $X_{\pm}^{B}$ and $Y_{\pm}^{A}$ to $Y_{\pm}^{B}$) to the connections ($X_{\pm}^{A}$ to $X_{\mp}^{B}$ and $Y_{\pm}^{A}$ to $Y_{\mp}^{B}$), we can extend the value of $\phi$ to [${\pi}/{2},\pi$] and [${3\pi}/{2},2\pi$] respectively. As such, $\phi$ can be realized through the whole range [$0,2\pi$]. Therefore, our result is succeed in emulating arbitrarily background magnetic flux and will open up the possibilities of fabricating nontrivial topological structures of high Chern numbers~\cite{27}.

\section{Conclusion}

In conclusion, we have theoretically realized the quantum spin Hall effect analog in the topological Lieb lattice constructed with radio-frequency linear circuits. To achieve this , we have properly designed a specific capacitor-inductor network to resemble the intrinsic SOCs, and have constructed the analog spin by mixing DOFs of voltages on sub-sites.
We have then investigated the spin-resolved topological edge mode and the topological phase transition of band structure varied with capacitances. They are characterized by the nonzero spin Chern numbers and corresponding sudden changes.
Finally, we have discussed the extension of $\pi/2$ phase change of hopping between sites to arbitrary phase values, which provides a new platform to design structure of high Chern number~\cite{27}.

Compared with previous methods of fabricating Lieb lattice, i.e., cold atoms, or optical waveguides by the direct laser writing technique~\cite{YL}, there is no doubt that the linear circuit is the more convenient and simple approach. Besides, the TLL has rich topological phases. As shown in Ref.~\cite{6}, the topological phase transition can be enriched by next-next-neighbor hopping, which can be readily realized in our circuit model just by introducing more capacitors to connect the inductor pairs. We believe that the study reported here may have potential applications in developing telecommunication, sensing and energy harvesting, engineering microwave topological metamaterials, and simulating strongly correlated states~\cite{17,strong}.

\acknowledgements
This work is supported by the National Key Research Program of China (No. 2016YFA0301101), the National Natural Science Foundation of China (Grant Nos. 11775159 and 61621001), and the National Youth 1000 Talents Program in China.

\begin{appendix}
\section{The equations of motion of the TLL circuit}
In this part, we show two examples to get the equations of motion summing in in Eqs.(\ref{eq5}). Based on the Kirchhoff's law, the total electric current from all direction to node $(m,n,X_{+})$ of $A$ is zero. As shown in Fig.\ref{fig1}(b), node $(m,n,X_+)$ of $A$ is connected with node $(m,n,X_-)$ of $A$ by a inductor with inductance $L$, and is connected with node $(m,n,X_+)$ of $B$, node $(m-1,n,X_+)$ of $B$, node $(m,n,X_+)$ of $C$ and node $(m,n-1,X_+)$ of $C$ by capacitors with capacitance $C_1$. So we get,
\begin{equation}\label{eqA1}
\begin{split}
  &\frac{V^{A}_{m,n,X_+}-V^{A}_{m,n,X_-}}{i\omega L}+i\omega C_1(V^{A}_{m,n,X_+}-V^{B}_{m,n,X_+})+i\omega C_1\\
  &(V^{A}_{m,n,X_+}-V^{B}_{m-1,n,X_+})+i\omega C_1(V^{A}_{m,n,X_+}-V^{C}_{m,n,X_+})   \\
  &+i\omega C_1(V^{A}_{m,n,X_+}-V^{C}_{m,n-1,X_+})=0.
\end{split}
\end{equation}
 The similar equation can be obtained for node $(m,n,X_{-})$ of $A$ which is connected with node $(m,n,X_+)$ of $A$ and is connected with node $(m,n,X_-)$ of $B$, node $(m-1,n,X_-)$ of $B$, node $(m,n,X_-)$ of $C$ and node $(m,n-1,X_-)$ of $C$ by capacitors with capacitance $C_1$:
\begin{equation}\label{eqA2}
\begin{split}
  &\frac{V^{A}_{m,n,X_-}-V^{A}_{m,n,X_+}}{i\omega L}+i\omega C_1(V^{A}_{m,n,X_-}-V^{B}_{m,n,X_-})+i\omega C_1 \\
  &(V^{A}_{m,n,X_-}-V^{B}_{m-1,n,X_-})+i\omega C_1(V^{A}_{m,n,X_-}-V^{C}_{m,n,X_-})   \\
  &+i\omega C_1(V^{A}_{m,n,X_-}-V^{C}_{m,n-1,X_-})=0.
\end{split}
\end{equation}
Using the definition $U_{m,n,i}^{j}=V_{m,n,i_+}^{j}-V_{m,n,i_-}^{j}$ with $i=X,Y$ and $j=A,B,C$, Eq.(\ref{eqA1}) minus Eq.(\ref{eqA2}) leads to
\begin{equation}\label{eqA3}
\begin{split}
 & U_{m,n,X}^{A}=\\
  &-\frac{\omega^{2}LC}{2}(-4t_1U_{m,n,X}^{A}+\\
  &t_1(U_{m,n,X}^{B}+U_{m-1,n,X}^{B}+U_{m,n,X}^{C}+U_{m,n-1,X}^{C})),
\end{split}
\end{equation}
which is the first equation in Eqs.(\ref{eq5}).

Next we show the equation of site $(m,n,X)$ of $B$. Like Eq.(\ref{eqA1}) and Eq.(\ref{eqA2}), the equation for node  $(m,n,X_+)$ and $(m,n,X_-)$ of $B$ can be obtained as,

\begin{equation}\label{eqA4}
\begin{split}
 &\frac{V_{m,n,X_+}^{B}-V_{m,n,X_-}^{B}}{i\omega L}+i\omega C_1(V_{m,n,X_+}^{B}-V_{m,n,X_+}^{A})+i\omega C_1\\
 &(V_{m,n,X_+}^{B}-V_{m+1,n,X_+}^{A})+i\omega C_2(V_{m,n,X_+}^{B}-V_{m,n,Y_-}^{C}) \\
 &+i\omega C_2(V_{m,n,X_+}^{B}-V_{m+1,n,Y_+}^{C})\\
 &+i\omega C_2(V_{m,n,X_+}^{B}-V_{m+1,n-1,Y_-}^{C})\\
 &+i\omega C_2(V_{m,n,X_+}^{B}-V_{m,n-1,Y_+}^{C})=0,
\end{split}
\end{equation}
and
\begin{equation}\label{eqA5}
\begin{split}
 &\frac{V_{m,n,X_-}^{B}-V_{m,n,X_+}^{B}}{i\omega L}+i\omega C_1(V_{m,n,X_-}^{B}-V_{m,n,X_-}^{A})+i\omega C_1\\
 &(V_{m,n,X_-}^{B}-V_{m+1,n,X_-}^{A})+i\omega C_2(V_{m,n,X_-}^{B}-V_{m,n,Y_+}^{C}) \\
 &+i\omega C_2(V_{m,n,X_=}^{B}-V_{m+1,n,Y_-}^{C})\\
 &+i\omega C_2(V_{m,n,X_-}^{B}-V_{m+1,n-1,Y_+}^{C})\\
 &+i\omega C_2(V_{m,n,X_-}^{B}-V_{m,n-1,Y_-}^{C})=0.
\end{split}
\end{equation}
Using Eq.(\ref{eqA4}) minus Eq.(\ref{eqA5}), we can get the third equation of Eqs.(\ref{eq5}).
\section{Calculation for band structure and the spin Chern number}
The band structure for the infinite TLL circuit model can be obtained from the matrix $M_{k}$ in Eq.(\ref{eq7}) using,
\begin{equation}\label{eqB1}
 \mathrm{Det}[M_{k}-\frac{1}{\omega^{2}}]=0,
\end{equation}
The topological properties of the bulk band are characterized by the spin Chern number $C_{spin}=\frac{C_{\uparrow}-C_{\downarrow}}{2}=C_{\uparrow}$\cite{spin}. The spin up Chern number for $n^{th}$ band is defined by,
\begin{equation}\label{eqB2}
  C_{n,\uparrow}=\frac{1}{2\pi i}\int_{BZ}d^{2}\mathbf{k}\Omega_{k_x,k_y,\uparrow}^{n}.
\end{equation}
$\Omega_{k_x,k_y,\uparrow}^{n}$ is the Berry curvature\cite{Berry} for spin up, which can be get through,

\begin{widetext}
\begin{equation}\label{eqB3}
 \Omega_{k_x,k_y,\uparrow}^{n}=\\i\sum_{n'\neq n}\frac{\langle U_{n,k,\uparrow}|\partial_{k_x}M_{k,\uparrow}| U_{n',k,\uparrow}\rangle \langle U_{n',k,\uparrow}|\partial_{k_y}M_{k,\uparrow}| U_{n,k,\uparrow}\rangle-(k_x\leftrightarrow k_y) }{(\frac{1}{\omega^{2}_{n,\uparrow}}-\frac{1}{\omega^{2}_{n',\uparrow}})^{2}}
\end{equation}
\end{widetext}
where $M_{k,\uparrow}$ is the sub-matrix of $M_{k}$ in Eqs.(\ref{eq7}) for the spin up block. $U_{n,k,\uparrow}$ is the eigenvector of the matrix $M_{k,\uparrow}$.
\section{Calculation for time evolution}
In this part, we show the time evolution equation for Fig.\ref{fig3}. The sample is composed of $10\times10$ unit cells, whose voltage difference is $U_{m,n,j}^{i}$ with $i=A,B,C$, $j=X,Y$ and $m,n=0,1\ldots,9$. For simplicify, we definite the voltage difference vector $\mathbf{U}=(\mathbf{U}_{X},\mathbf{U}_{Y})^{T}$, where $\mathbf{U}_{j}=(U_{0,j}^{A},U_{0,j}^{B},U_{0,j}^{C},\ldots,U_{N,j}^{A},U_{N,j}^{B},U_{N,j}^{C},\ldots)^{T}$ and the source vector $\mathbf{S}=(\mathbf{S}_{X},\mathbf{S}_{Y})^{T}$, where $\mathbf{S}_{j}=(S_{0,j}^{A},0,0,0,\ldots)^{T}$ with $N=10*n+m$. The equation of motion for $\mathbf{U}$ can be expressed as,
\begin{equation}\label{eqC1}
  \mathbf{\ddot{U}}+\mathbf{M}^{-1}\mathbf{U}=\mathbf{S}(t)
\end{equation}
where $\mathbf{M}$ is the coupling matrix which can be obtained from Eqs.(\ref{eq5}). For $S_{0,X}^{A}=e^{-\frac{(t-t_{0})^{2}}{\sigma^2}}\cos(\omega_{1}t)$ and $S_{0,Y}^{A}=e^{-\frac{(t-t_{0})^{2}}{\sigma^2}}\sin(\omega_{1}t)$ we get the spin up simulation in Fig.\ref{fig3}(a)(b)(c)(d). For $S_{0,X}^{A}=e^{-\frac{(t-t_{0})^{2}}{\sigma^2}}\cos(\omega_{1}t)$ and $S_{0,Y}^{A}=-e^{-\frac{(t-t_{0})^{2}}{\sigma^2}}\sin(\omega_{1}t)$ we get the spin down simulation in Fig.\ref{fig3}(e)(f)(g)(h). For $S_{0,X}^{A}=2e^{-\frac{(t-t_{0})^{2}}{\sigma^2}}\cos(\omega_{1}t)$ and $S_{0,Y}^{A}=0$ we get the results in Fig.\ref{fig3}(i)(j)(k)(l). For $S_{0,X}^{A}=2e^{-\frac{(t-t_{0})^{2}}{\sigma^2}}\cos(\omega_{2}t)$ and $S_{0,Y}^{A}=0$ we get the bulk results in Fig.\ref{fig3}(m)(n)(o)(p). The differential equation in Eq.(\ref{eqC1}) is solve with the fourth order Runge-Kutta method with the simulation step is less than ${1}/{100}$ of the periodic time $T={2\pi}/{\omega}$.
\section{Realization of arbitrary phase}
Here we show how the arbitrary phase values are realized in Fig.\ref{fig5}. Based on the Kirchhoff's law, the total electric currents from all direction to node $X_+$, $X_-$, $Y_+$ and $Y_-$ of $A$ are all zero, we thus obtain
\begin{equation}\label{eqD1}
\begin{split}
&\frac{V^{A}_{X_{+}}-V^{A}_{X_{-}}}{i\omega L}+i\omega C \cos(\phi)(V^{A}_{X_{+}}-V^{B}_{X_{+}})\\
&+i\omega C \sin(\phi)(V^{A}_{X_{+}}-V^{B}_{Y_{+}})=0,
\end{split}
\end{equation}
\begin{equation}\label{eqD2}
\begin{split}
&\frac{V^{A}_{X_{-}}-V^{A}_{X_{+}}}{i\omega L}+i\omega C \cos(\phi)(V^{A}_{X_{-}}-V^{B}_{X_{-}})\\
&+i\omega C \sin(\phi)(V^{A}_{X_{-}}-V^{B}_{Y_{-}})=0,
\end{split}
\end{equation}
\begin{equation}\label{eqD3}
\begin{split}
&\frac{V^{A}_{Y_{+}}-V^{A}_{Y_{-}}}{i\omega L}+i\omega C \cos(\phi)(V^{A}_{Y_{+}}-V^{B}_{Y_{+}})\\
&+i\omega C \sin(\phi)(V^{A}_{Y_{+}}-V^{B}_{X_{-}})=0,
\end{split}
\end{equation}
and
\begin{equation}\label{eqD4}
\begin{split}
&\frac{V^{A}_{Y_{-}}-V^{A}_{Y_{+}}}{i\omega L}+i\omega C \cos(\phi)(V^{A}_{Y_{-}}-V^{B}_{Y_{-}})\\
&+i\omega C \sin(\phi)(V^{A}_{Y_{-}}-V^{B}_{X_{+}})=0.
\end{split}
\end{equation}
Using definition $U_{i}^{j}=V_{i_+}^{j}-V_{i_-}^{j}$ with $i=X,Y$ and $j=A,B$, Eq.(\ref{eqD1}) minus Eq.(\ref{eqD2}) and Eq.(\ref{eqD3}) minus Eq.(\ref{eqD4}), we get two equations of motion for $U_{X}^{A}$ and $U_{Y}^{A}$ respectively,
\begin{equation}\label{eqD5}
\begin{split}
&U_{X}^{A}=-\frac{\omega^{2}LC}{2}((-\cos(\phi)-\sin(\phi))U_{X}^{A}\\
&+\cos(\phi)U_{X}^{B}+\sin(\phi)U_{Y}^{B}),
\end{split}
\end{equation}
and
\begin{equation}\label{eqD6}
\begin{split}
&U_{Y}^{A}=-\frac{\omega^{2}LC}{2}((-\cos(\phi)-\sin(\phi))U_{Y}^{A}\\
&+\cos(\phi)U_{Y}^{B}-\sin(\phi)U_{X}^{B}).
\end{split}
\end{equation}
Finally, by defining $U_{\uparrow}^{j}=U_{X}^{j}+iU_{Y}^{j}$ and $U_{\downarrow}^{j}=U_{X}^{j}-iU_{Y}^{j}$ with $j=A,B$, from Eq.(\ref{eqD5}) and Eq.(\ref{eqD6}), we obtain
\begin{equation}\label{eqD7}
  U_{\uparrow}^{A}=-\frac{\omega^{2}LC}{2}(-(\cos(\phi)+\sin(\phi)) U_{\uparrow}^{A}+e^{-i\phi}U_{\uparrow}^{B}),
\end{equation}
and
\begin{equation}\label{eqD8}
  U_{\downarrow}^{A}=-\frac{\omega^{2}LC}{2}(-(\cos(\phi)+\sin(\phi)) U_{\downarrow}^{A}+e^{i\phi}U_{\downarrow}^{B}).
\end{equation}
We can see that site $A$ and $B$ have an arbitrary phase coupling determined by the values of the capacitors. Utilizing the same scheme, by incorporating the interconnections into large size linear circuit network, we can realize topological nontrivial circuits.
\end{appendix}

\end{document}